\documentclass[prl,twocolumn,showpacs,superscriptaddress]{revtex4-1}
\usepackage{amsmath,graphicx}

\newcount\minute
\newcount\hour
\newcount\hourMins
\def\now
{
  \minute=\time
  \hour=\time \divide \hour by 60
  \hourMins=\hour \multiply\hourMins by 60
  \advance\minute by -\hourMins
  \zeroPadTwo{\the\hour}:\zeroPadTwo{\the\minute}
}\def\timestamp
{
  \today\ \now
}
\def\today
{
  \zeroPadTwo{\the\day}-\zeroPadTwo{\the\month}-\the\year%
}
\def\zeroPadTwo#1%
{
  \ifnum #1<10 0\fi
  #1%
}

\def\C {\scriptscriptstyle {\mathrm{C}}}
\def\sC {\scriptscriptstyle {\mathrm{SC}}}

\date{\timestamp}

\begin{document}

\title{Relaxation of weakly interacting electrons in one dimension}

\author{Zoran Ristivojevic}
\affiliation{Centre de Physique Th\'{e}orique, Ecole Polytechnique, CNRS, 91128 Palaiseau, France}
\affiliation{Laboratoire de Physique Th\'{e}orique--CNRS, Ecole Normale Sup\'{e}rieure, 24 rue Lhomond, 75005 Paris, France}
\affiliation{Materials Science Division, Argonne National Laboratory, Argonne, Illinois 60439, USA}

\author{K. A. Matveev}
\affiliation{Materials Science Division, Argonne National Laboratory, Argonne, Illinois 60439, USA}

\begin{abstract}
We consider the problem of relaxation in a one-dimensional system of interacting electrons. In the limit of weak interactions, we calculate the decay rate of a single-electron excitation, accounting for the nonlinear dispersion. The leading processes that determine the relaxation involve scattering of three particles. We elucidate how particular forms of Coulomb interaction, unscreened and screened, lead to different results for the decay rates and identify the dominant scattering processes responsible for relaxation of excitations of different energies. Interestingly, temperatures much smaller than the excitation energy strongly affect the rate. At higher temperatures the quasiparticle relaxes by exciting copropagating electron-hole pairs, whereas at lowest temperatures the relaxation proceeds via excitations of both copropagating and counterpropagating pairs.
\end{abstract}

\maketitle

Low energy excitations of a three-dimensional interacting electron system are fermionic quasiparticles that in many respects resemble bare electrons \cite{nozieresbook}. A quasiparticle excitation of energy $\varepsilon$ has finite decay rate $\tau^{-1}\propto \varepsilon^{2}$, where $\varepsilon$ is measured from the Fermi level. This fact is the foundation of the Fermi liquid theory and was confirmed experimentally by measuring the broadening of the Lorentzian-shaped spectral function \cite{valla+99PhysRevLett.83.2085}.

One-dimensional interacting fermions are conventionally described
within the exactly solvable Tomonaga-Luttinger model where particles
are assumed to have a linear dispersion. This model can be
diagonalized in terms of noninteracting bosonic excitations
\cite{Giamarchi,Haldane81} which have infinite lifetimes.  This
feature reveals an important limitation of the Tomonaga-Luttinger
model, because in general an excited system is expected to relax to
equilibrium. To study relaxation, one should therefore consider models
that take into account the curvature of the spectrum.  Recent
experimental observation of different equilibration rates of hot
electrons and holes in quantum wires \cite{barak+10} has confirmed the
importance nonlinear dispersion of electrons.  Study of interacting
electrons with nonlinear spectrum is a subject of intense theoretical
interest \cite{lunde+07PRB75, khodas+07PhysRevB.76.155402,pereira+09,karzig+10PhysRevLett.105.226407, matveev+10PhysRevLett.105.046401,micklitz+10PhysRevLett.106.196402, matveev+12PhysRevB.85.041102,lin+12}. The area of new physics beyond the Luttinger liquid formalism has been recently reviewed in Ref.~\cite{imambekov+12RevModPhys.84.1253}.

\begin{figure}
\includegraphics[width=1\columnwidth]{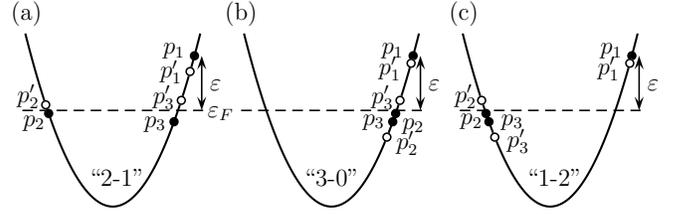}
\caption{A single electron excitation of energy $\varepsilon$ and momentum $p_1$ relaxes via scattering off two other electrons. Filled and empty circles represent, respectively, incoming and outgoing momenta. The one-dimensional topology of the Fermi surface determines the three cases (a), (b), and (c), which we refer to as the ``2-1''-, ``3-0''-, and ``1-2''-type processes. While at zero temperature only the ``2-1'' processes are allowed, all three types contribute to the quasiparticle decay at nonzero temperatures.}\label{fig1}
\end{figure}

In this paper we consider a system of spinless fermions and study the decay rate of a quasiparticle excitation placed above the Fermi level, see Fig.~\ref{fig1}. Since the complete study of the effects of nonlinear dispersion is very difficult, here we analyze the limit of weak interactions. In this case, the scattering processes can be classified by the number of colliding particles. Unlike in higher dimensions, two-particle processes do not lead to relaxation due to the conservation laws of energy and momentum. Therefore, the leading mechanism which provides finite relaxation rate involves scattering of three particles \cite{lunde+07PRB75}.

At zero temperature, the problem of relaxation was studied in Ref.~\cite{khodas+07PhysRevB.76.155402}, where the quasiparticle decay rate was found to behave as $\tau^{-1}\propto\varepsilon^8$. Here we study the effect of temperature $T$, and find dramatic departures from the $T=0$ case. We consider the situation when the temperature is much smaller than the energy of the excitation. Denoting the momentum of the excitation by $p_1$, we find the expression for its decay rate
\begin{align}\label{taudef}
\frac{1}{\tau}=\!\!\!\sum_{p_2,p_3\atop{p_1',p_2',p_3'}}\hspace{-0.3cm}'\  W_{123}^{1'2'3'} n_2n_3(1-{n}_{1'}) (1-{n}_{2'})(1-{n}_{3'}).
\end{align}
This equation accounts for scattering of the quasiparticle of momentum $p_1$ and two others with momenta $\{p_2,p_3\}$ into three outgoing states $\{p_1',p_2',p_3'\}$, see Fig.~\ref{fig1}. In Eq.~(\ref{taudef}) by $n_i=n_{p_i}$ we denote the Fermi occupation numbers, while $\Sigma'$ indicates summation over distinct states. The scattering rate $W_{123}^{1'2'3'}$ is determined by the Fermi golden rule expression, and reads
\begin{align}\label{rate}
W_{123}^{1'2'3'}=\frac{2\pi}{\hbar}|\mathcal{A}_{123}^{1'2'3'}|^2 \delta(E-E').
\end{align}
Here the $\delta$ function imposes conservation of the total energy, defined as $E=\varepsilon_{p_1}+\varepsilon_{p_2}+\varepsilon_{p_3}$ and similarly for the outgoing momenta. The three-particle scattering amplitude is defined as a vacuum expectation value
\begin{align}\label{amplitude-def}
\mathcal{A}_{123}^{1'2'3'}=\langle a_{p_1'}a_{p_2'}a_{p_3'}|V\frac{1}{E-H_0+i0^+}V| a^\dagger_{p_{1}}a^\dagger_{p_{2}}a^\dagger_{p_{3}} \rangle.
\end{align}
It is the central object that determines the relaxation rate (\ref{taudef}). The unperturbed Hamiltonian $H_0$ and the perturbation $V$ are taken in the form
\begin{align*}
H_0=\sum_p\varepsilon_p a_p^\dagger a_p,\quad V=\frac{1}{2L}\sum_{p_1,p_2,q}V_q a^\dagger_{p_1+q} a^\dagger_{p_2-q}a_{p_2}a_{p_1},
\end{align*}
where $a$ and $a^\dagger$ are the fermionic operators,
$\varepsilon_p=p^2/2m$, and $L$ is the system size. The two-body
interaction enters the Hamiltonian via its Fourier transform $V_q$.

The scattering amplitude (\ref{amplitude-def}) is very sensitive to the form of the two-body interaction $V(x)$. It vanishes for $V(x)\propto\delta(x)$, i.e., $V_q=\mathrm{const}$, corresponding to the contact interaction between spinless fermions. Nullification of the amplitude in this case arises because the Pauli principle prevents two electrons from sharing the same position. The amplitude (\ref{amplitude-def}) also vanishes for the Cheon-Shigehara model \cite{cheon+PhysRevLett.82.2536,imambekov+12RevModPhys.84.1253}, defined by $V(x)\propto \delta''(x)$, i.e., $V_q\propto q^2$. The latter belongs to the class of the so-called integrable models \cite{sutherland} for which there is no relaxation and therefore $\tau^{-1}=0$.

In quantum wires, the interaction between electrons is of a longer range. As an example of the most practical use, we consider the Coulomb interaction  defined by $V^{\C}(x)=e^2/|x|$, which has the Fourier transform $2e^2\ln(1/|p|)$. Interestingly, this logarithmic form gives a vanishing three-particle amplitude (\ref{amplitude-def}), although the model describing fermions interacting via Coulomb interaction is not expected to be integrable. It is therefore important to cut off the short distance singularity. This is done by accounting for the finite width of the wire $w$. One then obtains
\begin{align}\label{Vpcoulomb}
V_p^{\C}=2e^2\ln\left(\frac{\hbar}{|p|w}\right)\left(1+\frac{p^2 w^2}{\hbar^2}\right),
\end{align}
keeping the first two leading-order terms. Here and in the following
we neglect numerical factors under the logarithm because we consider
small momenta $|p|\ll \hbar/w$.

Calculation of the three particle amplitudes is rather tedious
\cite{supplement}. For unscreened Coulomb interaction
(\ref{Vpcoulomb}), one obtains the leading order result on the mass
shell
\begin{align}\label{ACfinal}
(\mathcal{A}_{123}^{1'2'3'})_{\C}=&- \frac{4e^4}{L^2} \frac{mw^2}{\hbar^2}\ln\left(\frac{\hbar}{p_r w}\right)\big[f_{\C}\left((\varphi'+\varphi)/2\right)\notag\\
&- f_{\C}\left((\varphi'-\varphi)/2\right)\big]\delta_{P,P'},
\end{align}
with the even periodic function $f_{\C}(\theta)=f_{\C}(\theta+{\pi}/{3})$ \begin{align}\label{fC}
&f_{\C}(\theta)=-\sum_{j=1}^3\frac{9\sin(\theta+2\pi j/3)\ln\left|\sin (\theta+2\pi j/3)\right|}{\sin(3\theta)}.
\end{align}
Instead of using the incoming momenta $\{p_j\}$ and the outgoing ones $\{p_j'\}$, for convenience here we have introduced Jacobi coordinates $P,p_r,\varphi$, defined as
\begin{gather}
\label{pi-Jacobi}
p_j=P/3+\sqrt{2/3}\,p_r\cos(\varphi-2\pi j/3),\quad j=1,2,3,
\end{gather}
and similarly for the outgoing momenta. Here $P=p_1+p_2+p_3$ has a simple meaning of the total momentum of the three colliding particles. The momentum $p_r$ is given by
\begin{gather}\label{relativemom}
p_r=\sqrt{(p_1-p_2)^2+(p_2-p_3)^2+(p_3-p_1)^2}/\sqrt{3},
\end{gather}
and measures the typical separation between the momenta. We note that the conservation laws impose $p_{r}'=p_r$.

We consider scattering at low temperatures, when all scattering particles should be in the vicinity of the two Fermi points. This enables us to classify particles as being right- or left-moving. Throughout this paper we study the decay of a right-moving excitation. The two other incoming particles can be classified in one of three ways: (i) one particle has positive momentum and the other negative, (ii) both have positive momenta, and (iii) both have negative momenta. These three configurations, respectively, have the values of the total momentum near $p_F$, $3p_F$, and $-p_F$, where $p_F$ is the Fermi momentum. Therefore, the incoming and the outgoing states must be in the same momentum configuration. Thus, we can distinguish between three different cases, which we call ``2-1'', ``3-0'', and ``1-2'' processes, see Fig.~\ref{fig1}. We choose notations where the momenta $p_j$ and $p_j'$ are always on the same branch of the Fermi surface, while the scattering amplitude takes into account all possible exchange processes.

At zero temperature, only processes of ``2-1'' type lead to relaxation \cite{khodas+07PhysRevB.76.155402}. On the contrary, at nonzero temperatures all three processes have nonzero rates. In the following we will calculate the rates and identify the dominant processes as the temperature increases.

We start our analysis by considering the ``2-1'' scattering process, Fig.~\ref{fig1}(a). In order to understand the energy dependence of the decay rate, let us consider the momentum change of the left-moving particle-hole pair. It is easily obtained from the conservation laws and reads $p_2'-p_2={(p_1'-p_1)(p_1'-p_3)}/{(p_3'-p_2)}$. Using the estimate $p_3'-p_2\approx 2p_F$, we can express the conservation laws contained in the scattering rate (\ref{rate}) as
\begin{align}\label{momchange}
\delta(E-E')\delta_{P,P'}\approx&\ \delta \bigg(p_2-p_2'-\frac{(p_1-p_1')(p_1'-p_3)}{2p_F}\bigg)\notag\\
&\times\frac{m}{2p_F}\delta_{p_1+p_2+p_3,p_1'+p_2'+p_3'},
\end{align}
where the momentum conservation comes from the amplitude (\ref{ACfinal}). We can now
employ the last equation to perform the summations over $p_2'$ and
$p_3'$ in the expression for the rate (\ref{taudef}). The summation
over the remaining three momenta $p_2,p_3$, and $p_1'$ then determines
the rate. From Eq.~(\ref{momchange}) we conclude that the typical
energy of the left-moving particle-hole pair is of the order of
$\varepsilon^2/\varepsilon_F$, where $\varepsilon\simeq v_F(p_1-p_F)$,
and $v_F$ is the Fermi velocity. Therefore, the integration over $p_2$
is restricted to an energy range of that width, while for both $p_3$ and $p_1'$ that range is of order $\varepsilon$. As a
result, at $T=0$ we find the phase space volume available for
scattering to be proportional to
$\varepsilon^2(\varepsilon^2/\varepsilon_F)$, where $\varepsilon_F$ denotes the Fermi energy \cite{khodas+07PhysRevB.76.155402}. Since the amplitude
(\ref{ACfinal}) depends on momenta only logarithmically, we infer the
scattering rate $\tau^{-1}\propto\varepsilon^4$. This result still
applies at very low temperatures $T\ll \varepsilon^2/\varepsilon_F$,
since then the thermal smearing of the occupation numbers is not
significant. However, a new behavior of the decay rate emerges in the
range of temperatures $\varepsilon^2/\varepsilon_F\ll T\ll
\varepsilon$, because the occupation numbers $n_2,n_{2'}$ in
Eq.~(\ref{taudef}) of the left-moving pair become thermally
smeared. As a result, the integration over momentum $p_2$ covers the
energy range of order $T$, and the phase space volume is proportional
to $\varepsilon^2 T$ \cite{imambekov+12RevModPhys.84.1253}. After careful calculation, using the amplitude (\ref{ACfinal}) and (\ref{fC}) for the unscreened Coulomb interaction, one finds
\begin{align}\label{tau21Cfinal}
\frac{1}{(\tau_\text{2-1})_{\C}}=&\left(\frac{e^2}{\hbar v_F}\right)^4 \left(\frac{p_F w}{\hbar}\right)^4\left(\ln\frac{\hbar}{p_Fw}\right)^2\notag\\&\times
\begin{cases}
c_1 {\varepsilon^4}/{\hbar\varepsilon_F^3}, & T\ll \varepsilon^2/\varepsilon_F \phantom{\frac{A^2}{B^2}}\\
c_2 {\varepsilon^2 T}/{\hbar\varepsilon_F^2},& \varepsilon^2/\varepsilon_F\ll T\ll \varepsilon \phantom{\frac{A^2}{B^2}}
\end{cases}
\end{align}
where the numerical prefactors are $c_1=(15-\pi^2)/1024\pi^3$ and $c_2=3(12-\pi^2)/64\pi^3$. The logarithm of the expression (\ref{tau21Cfinal}) originates from the first term in the amplitude (\ref{ACfinal}), using $p_r\sim p_F$. Compared to the decay rate $\tau^{-1}\propto\varepsilon^8$ of Ref.~\cite{khodas+07PhysRevB.76.155402}, our result (\ref{tau21Cfinal}) is significantly larger due to long-range nature of unscreened Coulomb interaction. On the other hand, the rate (\ref{tau21Cfinal}) is smaller than the one for electrons with spin \cite{karzig+10PhysRevLett.105.226407}, because interaction between spinless fermions is weaker as
a consequence of the Pauli principle.

At zero temperature the ``3-0'' processes, Fig.~\ref{fig1}(b), are not
allowed by the conservation laws
\cite{khodas+07PhysRevB.76.155402}. However, they do contribute to the
decay rate at finite temperatures. To evaluate the rate one can employ
a similar strategy to the one used above for the ``2-1'' processes. The momentum change of the initial excitation is $p_{1}-p_1'={(p_3'-p_3)(p_3'-p_2)}/{(p_{1}-p_2')}$, which enables us to express the conservation laws as
\begin{align}\label{momchange1}
\delta(E-E')\delta_{P,P'}\approx\ &  \delta\bigg(p_1-p_1'-\frac{(p_3'-p_3)(p_3'-p_2)}{p_1-p_F}\bigg)\notag\\
&\times\frac{m}{p_1-p_F}\delta_{p_1+p_2+p_3,p_1'+p_2'+p_3'}.
\end{align}
We can now use Eq.~(\ref{momchange1}) to perform the summation over the momenta $p_1'$ and $p_2$ in Eq.~(\ref{taudef}), which gives rise to a factor of $1/\varepsilon$ in the rate. The remaining summation over $p_2',p_3,p_3'$ is over the typical range of momenta of the order $T/v_F$ and delivers the factor $T^3$. The detailed calculation reveals the final result for the unscreened case
\begin{align}\label{tau30-high-C-main}
\frac{1}{(\tau_\text{3-0})_{\C}}=c_3\left(\frac{e^2}{\hbar v_F}\right)^4 \left(\frac{p_F w}{\hbar}\right)^4\left[\ln\frac{\hbar}{(p_1-p_F)w}\right]^2 \frac{T^3}{\hbar\varepsilon_F\varepsilon},
\end{align}
where $c_3\approx 1.13$ \cite{supplement}. The logarithmic prefactor arises from the amplitude (\ref{ACfinal}), using $p_r\sim p_1-p_F$.

The processes of ``1-2'' type are similar to the ``3-0'' ones. In
order to evaluate their decay rate, one can use the expression
for the momentum change $p_1-p_1'$ of the initial excitation of the same form as for the ``3-0'' case. However, in the present situation the denominator $p_1-p_2'$ should be replaced by $2p_F$, rather than by $p_1-p_F$. Therefore, the final result for the rate should be the same as in Eq.~(\ref{tau30-high-C-main}), provided one replaces $\varepsilon$ by $ 4\varepsilon_F$, which we confirmed by a careful calculation. Since $\varepsilon\ll\varepsilon_F$, the contribution of the``1-2'' processes is always subdominant.

We now turn to the case of the screened Coulomb interaction. We model the two-body potential as $V^{\sC}(x)=e^2/|x|-e^2/\sqrt{x^2+4d^2}$, where $d$ is the distance between the wire and a conducting plane representing nearby gates. At $|p|\ll \hbar/d\ll\hbar/w$, its Fourier transform is
\begin{align}\label{Vpscreenedcoulomb1}
V_p^{\sC}=2e^2\ln\left(\frac{d}{w}\right)-2e^2\frac{p^2d^2}{\hbar^2} \ln\left(\frac{\hbar}{|p|d}\right).
\end{align}
The first term in the right-hand side of Eq.~(\ref{Vpscreenedcoulomb1}) corresponds to the contact interaction $V\propto\delta(x)$, which does not affect spinless fermions. Therefore, the three-particle amplitude is determined by the second term of Eq.~(\ref{Vpscreenedcoulomb1}) and reads \cite{supplement}
\begin{align}\label{ASCfinal}
(\mathcal{A}_{123}^{1'2'3'})_{\sC}=&- \frac{8e^4}{3L^2} \frac{md^4p_r^2}{\hbar^4}\ln\left(\frac{\hbar}{p_rd}\right)[f_{\sC}((\varphi'+\varphi)/2)\notag\\ &- f_{\sC}((\varphi'-\varphi)/2))]\delta_{P,P'},
\end{align}
with the even periodic function $f_{\sC}(\theta)=f_{\sC}(\theta+{\pi}/{3})$
\begin{align}
f_{\sC}(\theta)=&-\sum_{j=1}^3\frac{9\left[5\sin(\theta+2\pi j/3)- \sin(5\theta-2\pi j/3)\right]}{4\sin(3\theta)}\notag\\ &\times\ln\left|\sin (\theta+2\pi j/3)\right|.
\end{align}
Compared to the unscreened interaction (\ref{Vpcoulomb}),
the screened interaction (\ref{Vpscreenedcoulomb1}) has two additional
powers of momentum. This is reflected in the corresponding amplitude (\ref{ASCfinal}), up to the logarithmic terms. As a result, the decay rates will have four additional powers of energy with respect to the unscreened case. For the ``2-1'' processes, the characteristic energy change is $\varepsilon$, which determines the decay rate at zero temperature $\tau^{-1}\propto \varepsilon^8$ \cite{khodas+07PhysRevB.76.155402}. After a careful calculation one obtains
\begin{align}\label{tau21SCfinal}
\frac{1}{(\tau_\text{2-1})_{\sC}}=&\left(\frac{e^2}{\hbar v_F}\right)^4 \left(\frac{p_F d}{\hbar}\right)^8 \left(\ln\frac{\hbar}{p_Fd}\right)^2\left(\ln\frac{\varepsilon}{\varepsilon_F}\right)^2 \notag\\& \times
\begin{cases}
c_4 \varepsilon^8/\hbar\varepsilon_F^7,& T\ll \varepsilon^2/\varepsilon_F \phantom{\frac{A^2}{B^2}}\\
c_5 \varepsilon^6 T/\hbar\varepsilon_F^6,& \varepsilon^2/\varepsilon_F\ll T\ll \varepsilon \phantom{\frac{A^2}{B^2}}\\
\end{cases}
\end{align}
where $c_4=15/32768\pi^3$ and $c_5=55/9216\pi^3$. For the ``3-0'' processes, the typical momentum change is $T/v_F$ and therefore one obtains four additional powers of temperature compared to Eq.~(\ref{tau30-high-C-main}),
\begin{align}\label{tau30-high-sC-main}
\frac{1}{(\tau_\text{3-0})_{\sC}}=&\ c_6\left(\frac{e^2}{\hbar v_F}\right)^4 \left(\frac{p_F d}{\hbar}\right)^8\left[\ln\frac{\hbar}{(p_1-p_F)d}\right]^2\notag\\ & \times\left(\ln\frac{\varepsilon}{T}\right)^2\frac{T^7}{\hbar\varepsilon_F^5\varepsilon},
\end{align}
where $c_6\approx 22.1$. The decay rates given by
Eqs.~(\ref{tau21Cfinal}) and (\ref{tau30-high-C-main}) for the Coulomb
interaction as well as Eqs.~(\ref{tau21SCfinal}) and
(\ref{tau30-high-sC-main}) for the screened Coulomb interaction are the main results of this paper. Now we comment on their physical meaning.

\begin{figure}
\includegraphics[width=0.95\columnwidth]{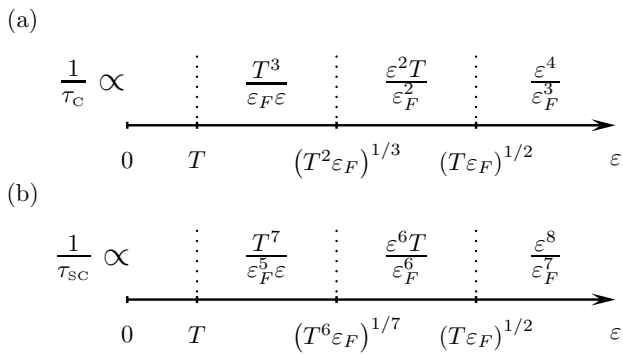}
\caption{Leading behavior of the decay rate of a quasiparticle of energy $\varepsilon$ for (a) unscreened Coulomb interaction (\ref{Vpcoulomb}), and for (b) screened Coulomb interaction (\ref{Vpscreenedcoulomb1}).}\label{fig2}
\end{figure}

The spectral function of a system of interacting electrons described
by the Tomonaga-Luttinger model displays a power-law edge singularity
on the mass shell
\cite{Giamarchi,imambekov+12RevModPhys.84.1253}. This divergence is a
signature of the infinite lifetime of excitations. However, once one
accounts for the curvature of the spectrum, the divergence disappears and the spectral function becomes broadened \cite{khodas+07PhysRevB.76.155402}. Therefore, the quasiparticles on the mass shell are subject to decay. The above-calculated decay rate describes broadening of the spectral function in the vicinity of the particle mass shell.  Our result (\ref{tau21SCfinal}) taken at zero temperature is consistent with Ref.~\cite{khodas+07PhysRevB.76.155402}.  It is worth noting,
however, that unlike our paper, where for the screened interaction all
the relevant momentum scales are assumed to be small compared with
$\hbar/d$, Ref.~\cite{khodas+07PhysRevB.76.155402} assumes
$p_F\gg \hbar/d$, i.e., the Fourier components of interaction
potential for momenta of the order of the Fermi momentum were
neglected.

The leading behavior of the decay rate is summarized in
Fig.~\ref{fig2}. Excitations of energies much larger than
$(T\varepsilon_F)^{1/2}$ decay with a temperature-independent rate.  For the unscreened interaction (\ref{Vpcoulomb}),
we infer a new energy scale
$\varepsilon^*\sim(T^2\varepsilon_F)^{1/3}$. Quasiparticles of
energies lower than $\varepsilon^*$ decay by exciting
co-propagating particle-hole pairs, while quasiparticles of energies
larger than $\varepsilon^*$ decay by exciting both co-propagating
and counter-propagating pairs. The same general picture applies in
the case of the screened Coulomb interaction (\ref{Vpscreenedcoulomb1}), but the crossover energy scale $\varepsilon^*$ is of order $(T^6\varepsilon_F)^{1/7}$. Interestingly, because the decay rate decreases with energy for the ``3-0'' processes but increases for the ``2-1'' ones, the rate has a minimum near the crossover energy scale $\varepsilon^*$.

To summarize, motivated by recent experiment \cite{barak+10} we have calculated the decay rate of quasiparticles in
weakly interacting one-dimensional electron systems. The dominant
mechanism of quasiparticle decay involves three electrons and is
illustrated in Figs.~\ref{fig1}(a) and \ref{fig1}(b).  The decay
rate shows nontrivial temperature dependence even at
$T\ll\varepsilon$, see Fig.~\ref{fig2}.

We acknowledge helpful discussions with L.~I.~Glazman, A.~Levchenko, and B.~I.~Shklovskii. This work at Ecole Normale Sup\'{e}rieure is supported by the ANR Grant No.~09-BLAN-0097-01/2 and at Argonne National Laboratory by the U.S. DOE, Office of Science, under Contract No. DE-AC02-06CH11357. Z.~R.~acknowledges financial support by Ecole Polytechnique.

\end{document}